\documentclass[12pt,fleqn]{article}
\usepackage{amssymb}
\usepackage{graphicx}
\usepackage{amsmath}
\usepackage{amsfonts}
\usepackage[latin1]{inputenc}

\begin{document}

\begin{centering}
{\leftskip=2in \rightskip=2in
{\large \bf D-particles and the localization limit in quantum gravity }}\\
\bigskip
\bigskip
\bigskip
\medskip
{\small {\bf Giovanni AMELINO-CAMELIA} and {\bf Luisa DOPLICHER}}\\
\bigskip
{\it Dipart.~Fisica Univ.~La Sapienza and Sez.~Roma1 INFN}\\
{\it P.le Moro 2, I-00185 Roma, Italy}

\end{centering}

\vspace{0.7cm}

\begin{center}
\textbf{ABSTRACT}
\end{center}

\baselineskip 11pt plus .5pt minus .5pt

{\leftskip=0.6in \rightskip=0.6in {\footnotesize Some recent
studies of the properties of D-particles suggest that in string
theory a rather conventional description of spacetime might be
available up to scales that are significantly smaller than the
Planck length. We test this expectation by analyzing the
localization of a space-time event marked by the collision of two
D-particles. We find that a spatial coordinate of the event can
indeed be determined with better-than-Planckian accuracy, at the
price of a rather large uncertainty in the time coordinate. We
then explore the implications of these results for the popular
quantum-gravity intuition which assigns to the Planck length the
role of absolute limit on localization. }}

\newpage 
\baselineskip 12pt plus .5pt minus .5pt
\pagenumbering{arabic} 

\setcounter{footnote}{0} \renewcommand{\thefootnote}{\alph{footnote}}

\pagestyle{plain}

\section{Introduction and summary}

The idea of an absolute limit on localization has a very long
tradition in quantum-gravity research~\cite{stachelearly}. Some
representative studies of various realizations of this idea can be
found in
Refs.~\cite{mead,padma,dopl1994,ahlu1994,ng1994,gacmpla,garay,casadio}.
As a result, much of the research in quantum gravity is guided by
the intuition that the localization of a spacetime event should be
fundamentally limited by (at least~\cite{ng1994,gacmpla}) the
Planck length, $L_P \sim 10^{-33}cm$, and that the length of a
spacetime interval could at best be measured with Planck-length
accuracy. The ``pre-D-brane'' perturbative string-theory
literature provides support for these expectations, since the
analysis of closed-string scattering indicates~\cite{venekonmen} a
corresponding role for the string length $L_s$ in measurability
limits (which is compatible with the Planck-length limit, since in
perturbative string theory $L_P < L_s$). However, some more recent
studies~\cite{dpart1,dpart2,dbrscatt1,dbrscatt2} have obtained
results in favour of the possibility that the collision region,
for collisions of D-particles, could have substringy and
subplanckian size. This is based on the
observation~\cite{dbrscatt2} that D-particles with appropriately
small relative velocity can be basically treated as ordinary
very-weakly-interacting point particles up to distances as small
as $g_s^{1/3} L_s$, without encountering comparatively large
relative-position quantum uncertainties, in a framework where $g_s
\ll 1$ and $L_P \sim g_s^{1/4} L_s$ (so that $g_s^{1/3} L_s \ll
L_P \ll L_s$).

These string-theory results have not affected the intuition of those working
in other areas of quantum-gravity research. The Planck length is still
assumed to set the absolute limit on localization and on length measurement.
The D-particle results are apparently perceived as some sort of peculiarity
of string theory, which should not affect the intuition of those approaching
the quantum-gravity problem from a different perspective. We intend to show
here that, on the contrary, the recent string-theory results on localization
are a manifestation of a more general weakness of the arguments that were
used to suggest that the Planck length sets the absolute limit on
localization and on length measurement.

We observe that the D-particle results on localization exploit the fact that
D-particles carry other charges in addition to the gravitational
charge/mass. The traditional Planck-scale-limit quantum-gravity intuition
was based on some analyses which implicitly assumed that the ``target
particle'' (the particle whose position is of interest) and the ``probe
particle'' (the particle used in the measurement procedure) interact only
gravitationally. The type of subplanckian accuracy achievable with
D-particles is instead the result of a combination of interactions. We
conclude that the result of subplanckian accuracy obtained with D-particles
exposes the fact that previous quantum-gravity analyses had assumed,
without justification, that the particles participating in the localization
procedure should only interact gravitationally.
Even in other approaches to the quantum-gravity
problem, possibly very different from string theory, it would not be
surprising to find an analogous result of subplanckian accuracy.
However we also find that the
example of D-particles suggests that one might be able to achieve
subplanckian accuracy in measurement of space position only at the cost of a
rather sizeable uncertainty on time measurement.

After some preliminaries (Section~2) on the Bohr-Rosenfeld approach to
measurability analysis in physics and on the standard Heisenberg microscope,
in Section~3 we analyze localization via a Heisenberg-microscope procedure
in the spirit of the quantum-gravity arguments~\cite{mead,padma} which
assume that the probe should be a massless neutral (but interacting
gravitationally) particle and lead to the traditional expectation of
separate Planck-length uncertainties for the measurement of space
coordinates, $\delta x \ge L_P $, and
time\footnote{We use conventions in
which $\hbar = c = 1$.}, $\delta t \ge L_P $, of an
event.

Examining this result we argue
that a possible cause of concern is the fact
that the associated quantum-gravity intuition
is based on measurement analyses all
assuming that a massless neutral probe should be used. While in
practice localization procedures typically do rely on massless (or anyway
relativistic) probes, in order to establish a fundamental measurability
limit one should consider the problem in very general terms.
It is possible that the probes that turn out to be useful for practical
reasons are not the ones conceptually best suited for the task of
localization. In order to establish a localization limit of more general
validity one should in particular consider both relativistic and
non-relativistic probes; moreover, one should consider both the case of
probes which only interact gravitationally (\textquotedblleft neutral
probes\textquotedblright ) and the case of probes which (in addition to
gravitational charge/mass) also carry other charges (\textquotedblleft
charged probes\textquotedblright ). In Section~4 we start by repeating the
same analysis of localization via a Heisenberg-microscope procedure, already
considered in Section~3, but replacing the massless neutral probe with a
nonrelativistic (massive) neutral probe. However, the analysis rapidly
suggests that the use of a nonrelativistic neutral probe cannot improve on
the localization limit established in the case of a relativistic neutral
probe. The end result is that measurements using a nonrelativistic neutral
probe can only achieve localization
at the level $\delta x\geq L_{P}/\sqrt{V_{P}^{3}}$, $\delta t\geq
L_{P}/\sqrt{V_{P}^{5}}$,
where $V_{P}$ is the velocity of the nonrelativistic probe ($V_{P}\ll 1$).

In Section~5 we consider the case of D-particles, as an example of probes
which do not interact exclusively gravitationally. Our analysis does not
require any of the most technical details of the string-theory framework
which supports the presence of D-particles. We only take into account the
fact that the underlying supersymmetry of the theoretical framework imposes
that D-particles, besides carrying gravitational charge, also carry another
charge associated (in an appropriate sense) to the gravitational charge
through supersymmetry. As a result of a compensation between these two
interactions~\cite{dbrscatt2} as long as the distance $d$ between two
D-particles is greater than $\sqrt{v} L_s$ (denoting with $v$ the relative
velocity and $L_s$ the string length) the energy stored in a
two-D-particle system can be described as $U \sim - L_s^6 v^4/d^7$
(up to an overall numerical factor of order 1).
This energy law replaces the corresponding
Newton energy law that applies to ``neutral'' particles. Another property of
D-particles which is used in our analysis is the fact that their
mass can be expressed as $g_s^{-1} L_s^{-1}$ in terms of
the string coupling and the string length.
Using these properties of D-particles
we find that the analysis
of a Heisenberg-microscope procedure of localization of a collision
between D-particles can indeed achieve
accuracy $\delta x \ge g_s^{1/3} L_s \sim g_s^{1/12} L_P$.
We also observe that such
a level of spatial localization can only be achieved at the cost of a rather
poor level of temporal localization. In fact, we find that the event is
temporally localized
with uncertainty $\delta t \sim g_s^{-1/3} L_s \sim g_s^{-7/12} L_P$
(and, since $g_s \ll 1$, this amounts to an uncertainty limit which
is significantly larger than the usually expected Planck-scale limit).

In Section~6 we stress that our observations for a localization measurement
procedure are also applicable (as one would expect) to procedures for the
measurement of the length of a spacetime interval.

The closing section (Section~7) is devoted to some comments on the
implications of our analysis for the intuition that should guide
quantum-gravity research. The example of D-particles suggests that
we should not necessarily assume that both $\delta x \ge L_P$
and $\delta t \ge L_P$ hold independently. Our findings however
provide support for the idea of a combined limit on position/time
measurement, something which could perhaps be schematized
with a relation of the type $\delta x \delta t \ge L_P^2$.

\section{Preliminaries}

\subsection{The Heisenberg microscope}

\label{s:micclass}

In Section~3, 4 and 5 we illustrate our point on localization in quantum
gravity within the familiar framework of the Heisenberg-microscope measurement
procedure. In order to render our discussion self-contained (and in order to
introduce some notation which will be useful in the following sections) we
provide here a brief review of the original (ordinary quantum mechanics,
neglecting gravity) formulation of the Heisenberg-microscope measurement
procedure~\cite{heisen, persico}. This original formulation of the
microscope is used to explore the implications of the uncertainty relations
for the accuracy with which it is possible to measure simultaneously the
position and the momentum of an electron. The position measurement is
achieved by means of a microscope, represented by a box with an opening
which allows incoming light to hit a photographic plate.
The electron is assumed to be located somewhere in the $xy$ plane
(see figure).
For simplicity we consider the
localization in the $x$ coordinate only. A photon scatters off the
electron via Compton scattering and is then collected in the
microscope. Were it possible to keep the electron fixed in the
same point of the $xy$ plane during the experiment, one could then
scatter a large number of photons off the electron, obtaining a
diffraction pattern on the photographic plate of the microscope.
The width of the first peak of the diffraction pattern would be
given by
\begin{equation}
l^{\prime }\simeq \frac{\lambda ^{\prime }}{\sin \beta }\simeq \frac{\lambda
^{\prime }}{\tan \beta }=\frac{\lambda ^{\prime }b}{L}\quad ,
\end{equation}%
where $\lambda ^{\prime }$ is the wavelength of the scattered photon,
and $\beta $, $L$ and $b$ are shown in the figure.
By establishing
the position of the center of the diffraction pattern one could then infer
the position of the electron in the $xy$ plane.
\begin{figure}[tbph]
\begin{center}
\includegraphics[width=8cm]{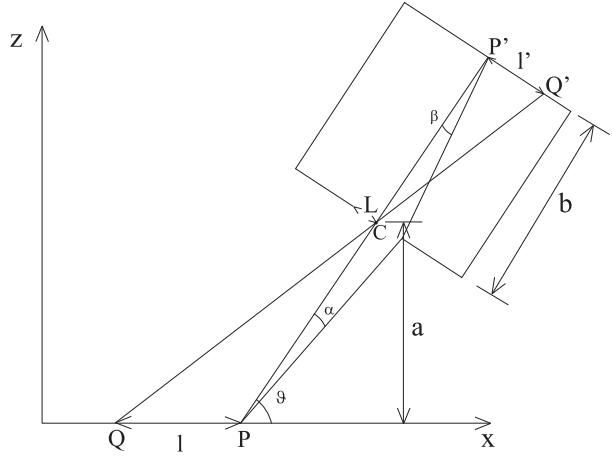}
\end{center}
\caption{In the Heisenberg-microscope setup,
if the target could be kept fixed
in $P$, one could scatter a large number of photons off
the target, obatining a diffraction pattern centered in $P^{\prime }$
on the photographic plate of the microscope.
If only one photon scatters off the target it is likely
that it reaches the photographic plate at some
point $Q^{\prime }$ within the first peak (of width $l^{\prime }$)
of the expected diffraction pattern.
Then $Q^{\prime }$ is most likely within a distance $l^{\prime }$
from $P^{\prime }$. The projection $Q$ of $Q^{\prime }$ through the
centre $C$ of the opening of the microscope is most likely within
a distance $l$ from $P$, which represents the actual position of
the target. $l$ is then
the uncertainty in the measurement of the position of the target.}
\label{fig:micsez}
\end{figure}
However, already the first
photon-electron collision would cause the electron to
(acquire some momentum and) move away from its
original position, which is the position that one is here attempting to
measure. Thus the position measurement must be based on a single point on
the photographic plate of the microscope, rather than on a whole diffraction
pattern. This leads to an uncertainty in the measurement of the
($x$-component of the) position of the electron which can be
estimated assuming
that the single point on the photographic plate of the microscope must have
appeared somewhere in the region which would have been occupied by the first
peak of the diffraction pattern. The uncertainty
is estimated as the width
of the peak projected on the $xy$ plane, and it is easy to verify
that
\begin{equation}
\delta x\gtrsim \frac{a}{b}l^{\prime }\simeq \frac{\lambda ^{\prime }a}{L}
~.
\label{jocDELTAx}
\end{equation}

In order to also estimate the uncertainty on the $x$-component of the
momentum of the electron, $p_{x}$, one can use the fact that in the
photon-electron collision a part of the initial momentum of the
photon is transferred to the electron, and after the collision $p_{x}$ is
given by
\begin{equation}
p_{x}\simeq p_{x}^{0}+\frac{1}{\lambda }-p_{x}^{\prime }\quad ,
\end{equation}%
where $p_{x}^{0}$ is the $x$-component momentum of the electron before the
scattering, and $p_{x}^{\prime }$ is the $x$-component of the momentum of
the photon after the scattering. The direction taken by the photon after the
scattering can be deduced from the fact that
it was collected by
the microscope. This allows
us to estimate the uncertainty on the $x$-component of the momentum of the
photon:
\begin{equation}
\delta p_{x}^{\prime }\sim \frac{1}{\lambda ^{\prime }}\sin \alpha \sin
\theta \gtrsim \frac{1}{\lambda ^{\prime }}\frac{L}{a}\quad ~,
\label{deltapinmic}
\end{equation}%
where $\alpha $ is shown in figure.

The uncertainty on the $x$-component of the momentum of the electron will be
of the same order:
\begin{equation}
\delta p_{x}\simeq \delta p_{x}^{\prime }\gtrsim \frac{1}{\lambda ^{\prime }}%
\frac{L}{a}\quad .
\end{equation}
Finally, combining this equation with equation (\ref{jocDELTAx})
one finds a limit on the simultaneous measurement
of the $x$-components of the position and momentum of the electron,
\begin{equation}
\delta x\delta p_{x}\gtrsim 1 ~,  \label{deltaxdeltap}
\end{equation}
consistently with
Heisenberg's uncertainty relation.
Of course, a similar conclusion can be reached for the $y$-
and $z$-directions, if needed, after a rotation of the apparatus.

The result (\ref{deltaxdeltap}) limits the simultaneous measurability of two
observables, $x$ and $p_{x}$, but does not constrain in any way the
measurability of a single one of them. In this framework there is no
in-principle reason (although it would not be feasible experimentally) that
prevents one from measuring $x$ exactly, $\delta x=0$, although this should
come at the cost of renouncing to all information on $p_{x}$.

In quantum gravity, as we will discuss in the following sections, it is
instead expected that there should be an in-principle obstruction for
achieving $\delta x = 0$.

\subsection{Bohr-Rosenfeld probes}

In the following sections we will analyze the Heisenberg-microscope
procedure from a quantum-gravity perspective. In the original Heisenberg
microscope one measures the position of an electron at rest and uses photons
as probes, but we find that a key issue from the quantum-gravity perspective
is the one of selecting the particles to be used as probes and the particles
whose position is to be determined via the Heisenberg-microscope procedure.
Clearly an electron at rest would not be a very sharp way to mark a
spacetime point in quantum gravity, since already in
relativistic quantum mechanics the localization of a particle of mass $M$
at rest is limited by its Compton wavelength, and therefore it appears that
large-mass particles should be preferable. On the other hand, as we shall
discuss in greater detail later, gravity introduces a source of uncertainty
that grows with the particle's mass. A balance of this competing sources of
uncertainty must be achieved
in order to obtain the true absolute limit on localization in quantum
gravity.

We want to stress here that it is not uncommon that the choice of the
particles used in the measurement procedure turns out to be a key point of
the analysis of a measurability limit. The best example is provided by
attempts (in the 1930s) to establish whether quantum electrodynamics sets a
measurability limit for the electromagnetic fields. Various arguments had
suggested that there might be a logical inconsistency in quantum
electrodynamics, since on the one hand the formalism predicts no absolute
limit on the measurability of the electromagnetic fields (if one is
willing to lose all information on some conjugate fields), whereas
the analysis of
several gedanken measurement procedures had provided evidence of an absolute
measurability limit. The situation was clarified in a study by Bohr and
Rosenfeld~\cite{bohrrose}, who proposed a gedanken measurement procedure
using probes of charge $Q$ and inertial mass $M$ in such a way to
obtain an uncertainty on the measurement of an electromagnetic field that is
proportional to the ratio $Q/M$. Considering the limit $Q/M \rightarrow 0$
the measurement procedure reproduced the result expected on the basis of the
formal structure of quantum electrodynamics, \textit{i.e.} an
uncertainty-free measurement of the relevant electromagnetic field.

In the Bohr-Rosenfeld analysis clearly the nature of the probes used in the
measurement procedure plays a key role, in light of the important dependence
on $Q/M$. This provides some guidance for the study we are here
reporting: we should be open to the possibility that the localization of a
spacetime point may depend significantly on the particles used in giving
operative meaning to that point.

The Bohr-Rosenfeld analysis also provides insight on the nature of the
particles to be used as probes in measurement analysis. In fact, it is
noteworthy that not only the measurability limit may be different if we use
different ``known particles'' (the ratio $Q/M$ has a different value for
electrons and muons), but it appears that we should consider all particles
that are allowed by the formalism, even when these particles have not been
observed in Nature. This is the line of analysis advocated
by Bohr and Rosenfeld when they contemplate the limit $Q/M \rightarrow 0$.
If one restricts the analysis to known particles it would of course not be
possible to gain access to the $Q/M \rightarrow 0$ limit and therefore the
(uncertainty-free) result expected on the basis of the formal structure of
quantum electrodynamics could not be achieved by the measurement procedure.
However, Bohr and Rosenfeld~\cite{bohrrose} stress that quantum
electrodynamics does not predict its constituents (\textit{e.g.} it predicts
how electrons interact but it does not predict the existence of electrons)
and its intrinsic structure should not be assumed to depend on elements
external to the theory, such as indeed the types of probes that happen to be
available in Nature. In actual measurements the measurability of an
electromagnetic field will be limited due to various practical facts,
including the fact that Nature does not make available to us particles with
arbitrarily small ratio $Q/M$, but quantum electrodynamics does predict in a
logically consistent way the absence of an in-principle limitation of the
measurability of electromagnetic fields, which finds its support in
measurement theory upon considering the limit of particles with arbitrarily
small ratio $Q/M$.

\section{Localization of an event marked by the collision between a photon
and a neutral massive particle}

\label{s:micfotneutr}

As mentioned, most approaches to quantum gravity are guided by the intuition
that the Planck length should set absolute limits on the measurability of
spatial distances, $\delta x \ge L_P$, and time intervals, $\delta t \ge L_P$
. The emergence of these measurability limits has been suggested by various
types of analyses~\cite{mead,padma,garay}, and can provide some guidance in
proposing schemes for spacetime
noncommutativity~\cite{dopl1994,ahlu1994,gacmaj} and certain types of spacetime
discretization~\cite{urrutia}. While in different measurement procedures the
details of the
analysis can be rather different, this result is basically inevitable if
quantum mechanics and general relativity are combined straightforwardly. In
fact, as we stressed earlier, quantum mechanics introduces
a ``Compton-wavelength uncertainty'' in localization, which decreases with
the particle mass, and for a particle of mass of order $L_{P}^{-1}$
leads to a localization uncertainty of order $L_{P}$. On the other hand
general relativity introduces a ``Schwarzschild-radius uncertainty'' in
localization, which increases with the particle mass, and for a particle of
mass of order $L_{P}^{-1}$ leads to a localization uncertainty which is also
of order $L_{P}$. So, as long as quantum mechanics and general relativity
are the only ingredients of the analysis, the absolute limit
on localization is necessarily of order $L_{P}$ (and is achieved in the
measurement of the position of particles with mass of order $L_{P}^{-1}$).

In order to illustrate the derivation of this limit within a complete
analysis of a localization measurement procedure, in this section we
consider, in the spirit of Refs.~\cite{mead,padma}, a reformulation of the
Heisenberg microscope gedanken experiment that takes into account
gravitational effects. For simplicity we describe the gravitational
interaction between the probe, a photon, and the target, a massive neutral
particle at rest, using a semi-Newtonian framework\footnote{%
As shown in Ref.~\cite{mead}, the estimates obtained using this
semi-Newtonian gravity turn out to be correct (using general relativity one
obtains the same estimates, after a somewhat more tedious analysis).}. In
Refs.~\cite{mead,padma,garay} the mass of the particle whose position is
being measured (which we will sometimes call ``target particle'') is not
specified, but of course, as we stressed above, its mass cannot be smaller
than $L_{P}^{-1}$ (otherwise its Compton wavelength would be larger than the
sought position accuracy $L_{P}$), and its mass cannot be larger
than $L_{P}^{-1}$ (otherwise the measurement procedure should bring
the photon inside the Schwarzschild
radius of the particle, since the probe-target distance must be at some
point of the order of $L_{P}$ if the position measurement procedure
must achieve $L_{P}$ accuracy). We will therefore implicitly assume
that the ``target particle'' is of Planckian mass.

The line of reasoning that we adopt here is slightly different from that of
the analysis of the Heisenberg-microscope procedure in
Subsection \ref{s:micclass}. While in ordinary quantum
mechanics (with its Galileo-Newton
spacetime background) one is exclusively interested in space-position
localization, from a quantum-gravity perspective one is of course interested
primarily in the spacetime localization of an event. Therefore our attention
is here shifted from the measurement of the position in space of a target
particle, to the measurement of the spacetime coordinates of the event of
collision between the probe and the target. Whereas in the original
Heisenberg-microscope analysis one considers the simultaneous measurement of
a coordinate and of the corresponding component of the momentum of the
target particle, here one is interested primarily in the measurement of two
coordinates, one space coordinate and the time coordinate of the spacetime
event of collision.

We must also stress that in principle the analysis of such a measurement
procedure from a quantum-gravity perspective should take into account a very
large number of potential sources of contributions to the overall
uncertainty. We do not claim to consider all of these possible sources of
uncertainty, but we focus on some which appear to be most significant for
the quantum gravity analysis. The uncertainties we do consider lead to an
absolute measurability limit. Other sources of uncertainty (that may be
present but are not considered in our analysis) could in principle lead to a
stricter limit, but the limit we obtain is absolute (cannot be violated,
since the presence of other sources of uncertainty of course
could not lead to an improved accuracy).

The possibility to lower/violate our absolute limit could be contemplated
however for other measurement procedures. The Heisenberg-microscope
procedure is one of the possible ways to localize a spacetime event and is
clearly affected by the absolute localization limit we describe. Although
this seems unlikely to us (and the analysis of a few alternative measurement
procedures will quickly lead the reader to an analogous intuition), one
cannot exclude the possibility that some other localization procedure may
not be affected by such a localization limit. We will take as working
assumption that this is not the case, but it cannot be excluded in principle.

We denote with $X$ the spatial coordinate and with $T$ the time
coordinate\footnote{These are the $X$ coordinate
and the time coordinate ``of the event''. The
careful reader might notice that the spacetime point marked by the
probe-target collision is not even sharply defined, since the probe and the
target have ``finite size" (intrinsic position uncertainty)
and there is no instant in the collision at which
the probe and the target have the same spacetime coordinates.
This, however,
is of merely academic concern: in a measurement procedure that aims for
space/time accuracies of order $L_P$ it is sufficient to introduce all
aspects of the analysis with corresponding accuracy, and indeed we will make
sure that the setup is such that the spacetime point marked by the collision
is specified with $L_P$ accuracy.} which are to be measured. As in the
traditional Heisenberg-microscope procedure the observable $X$ is here still
obtained by observing the position of arrival of the probe on the
photographic plate of the microscope, while $T$ can be obtained as
\begin{equation}
T\simeq t_{i}+X\quad ,
\end{equation}
where $t_{i}$ marks the instant when the experiment begins (the instant when the
probe is fired toward the target).

The uncertainties on the measure of $X$ and $T$, $\delta X$ and $\delta T$
respectively, can be described in this way\footnote{%
Our analysis looks only for an estimate of orders of magnitude, and
therefore we will discard all numerical factors of order 1.}:
\begin{equation}
\begin{split}
\delta X& \gtrsim \delta X_{c}+\delta x_{0}+d_{\min }+\delta x_{p}+\delta
x_{t} \\
\delta T& \gtrsim \delta t_{i}+\delta X
\end{split}
\label{incertezze}
\end{equation}%
where:

\noindent
$\delta X_{c}$ is the uncertainty due to the width of the opening of the
microscope;

\noindent
$\delta x_{0}$ is the uncertainty on the initial position of the probe
(which quantum mechanics relates to the uncertainty in the initial momentum
of the probe);

\noindent
$d_{\min }$ is the minimum distance between probe and target reached during
the measurement procedure;

\noindent
$\delta x_{p}$ is the uncertainty on the position of the probe at the moment
of the collision;

\noindent
$\delta x_{t}$ is the uncertainty on the position of the target at the
moment of the collision;

\noindent
$\delta t_{i}$ is the uncertainty on the instant when the probe is fired
toward the target.

We start by noticing that $\delta X_{c}$ is an uncertainty
of classical-mechanics origin, and it can be reduced at will
by varying the width of the opening of the microscope.
Therefore $\delta X_{c}$ cannot be significant in establishing
an absolute limit on localization.

Concerning $\delta x_{t}$ we must stress that clearly (on the basis of
our considerations concerning the competing contributions to
the $\delta x_{t}$ uncertainty due to the ``Compton-wavelength uncertainty''
and the ``Schwarzschild-radius uncertainty'')
one must necessarily find $\delta x_{t} \ge L_P$.
Therefore the best we can hope for is a localization at the
level $\delta X \sim L_P$. We intend
to show, through an analysis of the other sources of uncertainty,
that $\delta X \sim L_P$ can be achieved. We will in general denote
with $\delta x^{\ast }$ the accuracy for which one aims in the measurement
procedure. In this case, with $\delta x^{\ast } \sim L_P$, consistently we
must take
\begin{equation}
\begin{split}
\delta x_{0}& \lesssim \delta x^{\ast }\sim L_{P} \\
d_{\min }& \lesssim \delta x^{\ast }\sim L_{P}
\end{split}
\label{richieste}
\end{equation}

The fact that, as stressed above, our analysis assumes that the mass of the
target is of the order of the Planck scale, $m_{t} \sim L_P^{-1}$, is
compatible with the choice $d_{\min } \sim L_{P}$ (we remind the reader that
the minimum probe-target distance $d_{\min }$ must not be smaller than the
Schwarzschild radius of the target particle, $d_{\min }\gtrsim Gm_{t}$).

From the fact that $\delta x_{0} \lesssim L_{P}$ it follows that
\begin{equation}
\delta p_{p}^{0} \sim \frac{1}{\delta x_{0}} \gtrsim \frac{1}{L_P}
\end{equation}%
where $p_{p}^{0}$ is the initial momentum of the probe, and $\delta
p_{p}^{0} $ is the corresponding uncertainty. From the fact that of course
we must demand that $p_{p}^{0}{\gtrsim }\delta p_{p}^{0}$, one finds that
\begin{equation}
p_{p}^{0}{\gtrsim }\delta p_{p}^{0}\sim \delta x_{0}^{-1}\gtrsim E_{P}\quad ,
\label{gammaenergy}
\end{equation}%
where $E_{P}$ is the Planck energy.

Next we should consider the contribution to the uncertainties that
originates from the uncertainty in the energy stored
in the probe-target system in the course of the collision process. For our
purposes (since we are only looking for an order-of-magnitude estimate) it
is sufficient to consider this issue only for the stage of the collision in
which the probe-target distance is of order $d_{\min }$.
At such short probe-target distances there is a rather
strong gravitational field, which is significantly affected by the
uncertainty $\delta p_{p}^{0}$. However, this strong gravitational field is
only present when indeed the probe-target distance is small. At early and
late times in the measurement procedure nearly all the energy of the system
is stored as kinetic energy, and we can expect that the kinetic energy of
the probe at early and late times will be of the same
order\footnote{In the
collision the target (which was initially at rest) will only take away from
the probe a small (negligible) fraction of the kinetic energy, even though
the target acquires a nonnegligible momentum. This is due to
to the large mass of the target (as compared to the massless particle
used as probe). In general, for a photon scattered along the direction
characterized by scattering angle $\theta$ the relation between
the momentum of the probe before the collision, $p_p$, and the momentum
of the probe after the collision, $p^{\prime}_p$, is
set by the formula $p^{\prime }_p = p_p /[1+p_p ( 1-\cos \theta )/m_t]$,
where $m_t$ denotes again the mass of the ``target" particle.}.

In order to estimate $\left( \delta p_{p}\right) _{x}$, the uncertainty on
the $x$-component of the probe's momentum, we can proceed as in the
preceding section, relying on the observation that the measurement procedure
requires that the probe reaches the photographic plate of the microscope.
Therefore
\begin{equation}
\left( \delta p_{p}\right) _{x}\lesssim
p_{p}^{\prime }\sin \alpha \simeq
p_{p}^{\prime }\tan \alpha
=p_{p}^{\prime }\frac{L}{a}\sim p_{p}\frac{L}{a} \quad ,
\label{dpinmic}
\end{equation}
where $p_{p}^{\prime }$ is the momentum of the probe after
the scattering, $\alpha $ is defined in the figure,
and we also used the observation that $p_{p}^{\prime }\sim p_{p}$.

From this it follows that
\begin{equation}
\delta x_{p}\gtrsim \frac{1}{p_{p}\frac{L}{a}}\quad .  \label{dxinmic}
\end{equation}
This $\delta x_{p}$ describes the uncertainty on the probe's $x$ coordinate
at the moment in which the probe reaches the opening of the microscope. This
same $\delta x_{p}$ represents a good estimate of the uncertainty on the
probe's $x$ coordinate at the moment of the collision (all distance scales
in the measurement procedure are certainly small enough that the spread of
the probe's wave packet is negligible~\cite{gold}).

Finally we need an estimate for $\delta x_{t}$, the uncertainty on the
target's $x$ coordinate at the moment of the collision. A key contribution
to this uncertainty (in addition to the Compton-wavelength uncertainty,
which is under control with our choice of target-particle mass) originates
from the gravitational interaction between probe and target. By momentum
conservation we have that the gravitational interaction must induce
correlated variations of the target's momentum and of
the probe's momentum,
\begin{equation}
\Delta p_{t}\left( t\right) \simeq \Delta p_{p}\left( t\right) \simeq \frac{%
Gp_{p}m_{t}}{d\left( t\right) }\quad ,
\end{equation}
so that
\begin{equation}
\left( \delta v_{t}\right) _{x}\left( t\right) =\frac{\left( \delta
p_{t}\right) _{x}\left( t\right) }{m_{t}}\simeq \frac{\Delta p_{p}\left(
t\right) }{m_{t}}\frac{L}{a}\quad
\end{equation}
(Note that in this equation we also used the fact that, because of
the large mass of the target, while the target's momentum is nonnegligible,
its velocity $v_{t}$ is small.)

Our estimate of $\delta x_{t}$ is based on the fact that it
must be greater than (but comparable to) the corresponding
uncertainty that develops in a small time interval, of size $d_{\min }$,
around $T$:
\begin{align}
\delta x_{t}& \geq \int_{T-d_{\min }}^{T}|(\delta v_{t})_{x}|dt\simeq
\int_{T-d_{\min }}^{T}\left\vert \frac{\Delta p_{p}\left( t\right) }{m_{t}}
\right\vert \frac{L}{a}dt\sim \left. \frac{\Delta p_{p}\left( t\right) }{
m_{t}}\right\vert _{d_{\min }}d_{\min }\frac{L}{a}
\simeq Gp_{p}\frac{L}{a}\quad .
\end{align}

We are finally ready to combine all the contributions to $\delta X$ listed
in (\ref{incertezze}), obtaining
\begin{equation}
\delta X\gtrsim \delta X_{c}+\delta x_{0}+d_{\min }
+\delta x_{p}+\delta x_{t}\gtrsim \delta X_{c}
+\delta x_{0}+d_{\min }+\frac{1}{p_{p}\frac{L}{a}}
+Gp_{p}\frac{L}{a}\quad .
\end{equation}
Not all of these contributions are equally significant from our
quantum-gravity perspective. We already stressed that $\delta X_{c}$
depends on the structure of the microscope, and can be reduced at will,
while $\delta x_{0}$ and $d_{\min }$ depend on
the characteristics of the probe and the target, and are chosen
precisely in order to reach the intended $\delta x^{\ast } \sim L_P$
accuracy goal. One easily sees that, in order to verify that this sought
accuracy can be actually reached, it is necessary to examine
the contributions $\delta x_{p}$ and $\delta x_{t}$,
since one of them decreases as $p_{p}$ is increased
while the other
increases\footnote{This is a key point: in ordinary
quantum mechanics one achieves as good a
localization as desired by increasing the momentum of the probe, whereas by
taking into account gravitational effects an increase in the momentum of the
probe is not always beneficial for localization. A higher frequency photon,
while resolving smaller lengths, brings about a more intense gravitational
field, which ends up introducing a bigger uncertainty in the target's
position.} with $p_{p}$ (and therefore they cannot be both made small at
will). The minimum-uncertainty case corresponds to
\begin{equation}
\min \left( \delta x_{p}+\delta x_{t}\right) =\sqrt{\frac{1}{p_{p}\frac{L}{a}%
}Gp_{p}\frac{L}{a}}=L_{P}\quad .
\end{equation}
Thus, we find (as expected) that it is indeed possible to reach the intended
accuracy of the order of the Planck length:
\begin{equation}
min (\delta X ) \sim L_{P}\quad .  \label{dxgeqlp}
\end{equation}
The analysis of the $\delta x_{p}$ and $\delta x_{t}$ contributions also
reassures us that indeed it would have not been possible to aim for anything
better than Planck-length accuracy: even choosing $\delta x_{0}$
and $d_{\min }$ smaller than the Planck length one would still
inevitably find a
Planck-length uncertainty due to the combined contribution
from $\delta x_{p}$ and $\delta x_{t}$.

For the uncertainty on $T$ we find
\begin{equation}
\delta T\gtrsim \delta t_{i}+\delta X\gtrsim L_{P}\quad ,  \label{dtgeqlp}
\end{equation}
where
\begin{equation}
\delta t_{i}\sim \frac{1}{\delta E_{p}}\sim \frac{1}{\delta p_{p}}\sim
\delta x_{0}\sim L_{P}\quad .
\end{equation}

The result (\ref{dxgeqlp}) for $\delta X$ is widely
accepted~\cite{mead,padma, garay}, and can be obtained
on the basis of several types of
measurement analyses, of which our Heisenberg-microscope procedure is only
one example. Studies of the limit on $\delta T$ are somewhat less numerous,
but our result (\ref{dtgeqlp}) also finds support in all the related
literature~\cite{mead, padma, garay}.

As mentioned, the intuition emerging from these analyses is also consistent
with some results obtained in the ``pre-D-brane'' perturbative string-theory
literature, since the analysis of closed-string scattering~\cite{venekonmen}
provides support for $\delta X \ge L_s$ (where $L_s$ is the string length,
and in the relevant framework $L_P < L_s$).

\section{Localization of an event marked by the collision between two
neutral massive particles}

\label{s:micneutr}

The localization limit $\delta X \ge L_P$, $\delta T \ge L_P$ is widely
accepted within the quantum-gravity community. We feel that it is often
overlooked that the derivation of this localization limit relies on two
crucial ingredients: the nature and strength of the gravitational
interactions and the fact that a massless particle with energy uncertainty $%
\delta E$ has position uncertainty $1/\delta E$. As stressed earlier, while
in practice localization procedures typically do rely on massless (or anyway
relativistic) probes, according to the Bohr-Rosenfeld line of analysis~\cite%
{bohrrose} it is necessary to wonder whether the probes that turn out to be
useful for practical reasons are the ones conceptually best suited for the
task of localization. In order to establish a localization limit of more
general validity one should in particular consider both relativistic and
non-relativistic probes; moreover, one should consider both the case of
probes which only interact gravitationally (``neutral probes'') and probes
which, in addition to gravitational charge/mass, also carry other charges
(``charged probes''). In this Section we start by repeating the same
analysis of localization via a Heisenberg-microscope procedure, already
considered in the previous Section, but replacing the massless neutral probe
with a nonrelativistic (massive) neutral probe.

We denote by $m_{p}$ and $m_{t}$ respectively the masses of the probe and
the target. Our analysis can
follow the same steps already discussed in the previous section. The only
key differences originate from the fact that here the speed of the probe is
taken to be much smaller than 1. Again the $X$ coordinate of the collision
event can be obtained by observing the position of arrival of the probe on
the photographic plate of the microscope, while $T$ can be here estimated
from
\begin{equation}
T\simeq t_{i}+\frac{X}{v_{p}}\quad ,
\end{equation}%
where $v_{p}$ is the speed of the probe.

$\delta X$ and $\delta T$ are again a combination of various contributions:
\begin{equation}
\begin{split}
\delta X& \gtrsim \delta x_{0}+d_{\min }+\delta x_{p}+\delta
x_{t} \\
\delta T& \gtrsim \delta t_{i}+\frac{\delta X}{v_{p}}+\frac{X}{v_{p}^{2}}%
\delta v_{p}\quad ,
\end{split}
\label{incertezze2}
\end{equation}%
where $\delta x_{0}$, $d_{\min }$, $\delta x_{p}$, $\delta
x_{t}$, $\delta t_{i}$ have been defined in the preceding Section~\ref%
{s:micfotneutr}, while $\delta v_{p}$ is the uncertainty on $v_{p}$.
(Of course, also in this case there is a $\delta X_{c}$
contribution, which we are omitting since, as clarified in the previous
section, it is of classical-mechanics origin and can therefore be reduced
at will.)

Our objective here is to show that, using a nonrelativistic probe,
one finds an absolute limit on
localization with minimum uncertainty
larger than $L_P$. We therefore set up the measurement procedure aiming
for Planck-length accuracy
\begin{align}
\delta x_{0}& \lesssim \delta x^{\ast }\sim L_{P} \\
d_{\min }& \lesssim \delta x^{\ast }\sim L_{P}\quad .  \label{dminmin}
\end{align}
and show that there are some contributions to $\delta X$ and $\delta T$
which lead to a worse-than-Planckian result.

Again we observe that from (\ref{dminmin}) one obtains
\begin{equation}
\delta p_{p}^{0} \sim \frac{1}{\delta x_{0}} \gtrsim \frac{1}{L_P}
\end{equation}
and that from the requirement $p_{p}^{0}{\gtrsim }\delta p_{p}^{0}$
it then follows that
\begin{equation}
p_{p}^{0}{\gtrsim }\delta p_{p}^{0}\sim \delta x_{0}^{-1}\gtrsim E_{P}\quad ,
\label{gammaenergybis}
\end{equation}
(we are again denoting with $p_{p}^{0}$ the initial momentum of the probe
and with $\delta p_{p}^{0}$ the corresponding uncertainty).

Here, with the nonrelativistic probe, the fact that $\delta p_{p}^{0}
\gtrsim 1/L_P$ implies $\delta v_{p}\gtrsim 1/(m_p L_P) $, a rather large
velocity uncertainty. Since we will anyway find that the use of the
nonrelativistic probe
leads to worse-than-Planckian localization, we allow ourself
a ``rather optimistic'' attitude\footnote{As mentioned, in order
to keep the ``Schwarzschild-radius uncertainty'' at
or below the accuracy goal $L_P$, it is necessary to
assume $L_P^{-1} \gtrsim m_p$. On the other hand one
must have $v_{p} < 1$ in order to be
consistent with the nonrelativistic nature of the probe and
one must have $v_{p} \ge \delta v_{p} \gtrsim 1/(m_p L_P) $
in order to be able to aim the probe toward the target.
The fact that this requirements cannot be
simultaneously implemented is already an indication of the fact
that the $L_P $ accuracy is not within the reach of a measurement
procedure using a nonrelativistic neutral probe. We will set this
concern aside, and find that it is anyway impossible to
achieve $L_P$ accuracy using a nonrelativistic
neutral probe.} concerning the possibility
of satisfying the requirements $v_{p}\gtrsim \delta v_{p}$
and $v_{p} < 1$.

In order to estimate $\delta x_{t}$ we must consider the effects induced by
the gravitational probe-target interaction at least over a small
time interval, of size $d_{\min }$, around $T$.
We observe that also in this case it is legitimate to
assume $p_{p}^{\prime }\sim p_{p}$, \textit{i.e.} the momentum
of the probe at early
and late times is of the same order\footnote{This
can be easily verified by looking at
the formulae that describe nonrelativistic scattering between
particles of equal mass, in the case where one of the particles
is initially at rest while the other particle initially carries
a large momentum. For generic scattering angle one finds that
the two particles carry final momenta of the same order of
magnitude (which of course is the same order of magnitude of
the initial momentum of the probe). One well-known exception is
the case of a ``central collision", in which the particle initially
at rest ends up carrying all the momentum in the final configuration,
but of course this is not a viable
option for the setup of our microscope.}.
This allows us to proceed just like
in Eqs.~(\ref{dpinmic})-(\ref{dxinmic}) of the previous section, finding
\begin{equation}
\delta x_{p}\gtrsim \frac{1}{p_{p}\frac{L}{a}}\quad .
\end{equation}
and finding for $\delta x_{t}$ (again proceeding as in the previous section)
\begin{equation}
\delta x_{t}\geq \int_{T-\frac{d_{\min }}{v_{p}}}^{T}|(\delta
v_{t})_{x}|dt\simeq \int_{T-\frac{d_{\min }}{v_{p}}}^{T}\left\vert \frac{%
\Delta p_{p}\left( t\right) }{m_{t}}\right\vert \frac{L}{a}dt\sim \left.
\frac{\Delta p_{p}\left( t\right) }{m_{t}}\right\vert _{d_{\min }}\frac{%
d_{\min }}{v_{p}}\frac{L}{a}\quad .
\label{joc77}
\end{equation}
where $\Delta p_{p}(t)$ is the variation of the momentum of the probe
induced by the gravitational interaction at the instant $t$ (where $t$ is
taken to be close to $T$).

To estimate $\Delta p_{p}(t)$ we can argue as follows.
Considering the probe's
total energy $E_{p}$ and the gravitational potential
energy $U\left(t\right) $ originated by the interaction
with the target
\begin{equation}
E_{p}=m_{p}+\frac{p_{p}^{2}}{2m_{p}}+U\left( t\right) \quad ,
\end{equation}
we can estimate the probe's momentum as
\begin{equation}
p_{p}=\sqrt{2m_{p}\left( E_{p}-U\left( t\right) -m_{p}\right) }\quad ,
\end{equation}
and we find that
\begin{equation}
\Delta p_{p}\simeq -\frac{m_{p}}{p_{p}}\Delta U\sim \frac{m_{p}}{p_{p}}\frac{%
Gm_{t}m_{p}}{d_{\min }}=\frac{Gm_{t}m_{p}}{v_{p}d_{\min }}\quad .
\end{equation}
Using (\ref{joc77}) this allows to derive
\begin{equation}
\delta x_{t}\sim \left. \frac{\Delta p_{p}\left( t\right) }{m_{t}}%
\right\vert _{d_{\min }}\frac{d_{\min }}{v_{p}}\frac{L}{a}\simeq \frac{Gm_{p}%
}{v_{p}d_{\min }}\frac{d_{\min }}{v_{p}}\frac{L}{a}=\frac{Gp_{p}}{v_{p}^{3}}%
\frac{L}{a}\quad ,
\end{equation}
and finally, combining all these results, we find that the total uncertainty
on $X$ is estimated by
\begin{equation}
\delta X\gtrsim \delta x_{0}+d_{\min }+\delta x_{p}
+\delta x_{t}\gtrsim \delta x_{0}+d_{\min }
+\frac{1}{p_{p}\frac{L}{a}}
+ \frac{Gp_{p}}{v_{p}^{3}}\frac{L}{a}\geq \frac{1}{p_{p}\frac{L}{a}}
+\frac{Gp_{p}}{v_{p}^{3}}\frac{L}{a}\quad ,
\end{equation}%
where on the right-hand side we are inviting the reader to focus on two of
the contributions. Those two contributions
are sufficient for establishing our result that using
the nonrelativistic neutral probe only a worse-than-Planckian accuracy is
achievable. In fact,
\begin{equation}
\delta X\geq \min \left( \delta x_{p}+\delta x_{t}\right)
=\sqrt{\frac{1}{p_{p}\frac{L}{a}}\frac{Gp_{p}}{v_{p}^{3}}\frac{L}{a}}
=\frac{L_{P}}{v_{p}^{\frac{3}{2}}}\quad ,
\end{equation}
which indeed represents worse-than-Planckian accuracy (since the hypothesis
of using a nonrelativistic probe requires $v_{p}<1$).

The use of a nonrelatistic neutral probe is even more obviously costly for
the accuracy in the measurement of $T$. In fact, on the basis of the
observations reported above one finds
\begin{equation}
\delta T\gtrsim \delta t_{i}+\frac{\delta X}{v_{p}}+\frac{X}{v_{p}^{2}}%
\delta v_{p}\sim \delta t_{i}+\frac{\delta X}{v_{p}}+Xv_{p}^{-\frac{1}{2}%
}\quad ,
\end{equation}%
with
\begin{equation}
\delta t_{i}\sim \frac{1}{\delta E_{p}}\sim \frac{1}{v_{p}\delta p_{p}}\sim
\frac{\delta x_{0}}{v_{p}}\quad .
\end{equation}

Since in a reasonable Heisenberg-microscope setup the distance $X$ (distance
between the point where the experimenter introduces the probe and the point
where the probe-target collision occurs) is macroscopic, the contribution $%
Xv_{p}^{-1/2}$ can be very large. And in any case the contribution $\delta
X/v_{p}$ implies that $\delta T$ is clearly larger than $L_{P}$: $\delta X$
is already larger than $L_{P}$ by at least $v_{p}^{-3/2}$ and therefore $%
\delta T$ is larger than $L_{P}$ by at least $v_{p}^{-5/2}$.

\section{Localization of an event marked by the collision between two
D-particles}

\label{s:micDpar}

In our investigation of the expectation that a localization limit $\delta X
\ge L_P$, $\delta T \ge L_P$ should hold in quantum gravity we observed that
this limit is essentially based on the analysis of measurement procedures in
which a massless neutral particle probes the position of a
massive target particle, and we argued that a more robust estimate of the
localization limit could be achieved by considering both relativistic and
non-relativistic probes, and by considering both the case of neutral probes
(probes which only interact gravitationally) and the case of charged probes
(probes which, in addition to a gravitational charge/mass, also carry other
charges). In the previous section we showed that replacing the massless
neutral probe with a nonrelativistic neutral probe one cannot improve on the
expected localization limit $\delta X \ge L_P$, $\delta T \ge L_P$. In this
Section we use the case of D-particles as an illustrative example of the
role that charged particles might have in allowing to achieve an improved
level of localization.

D-particles are zero-dimensional (pointlike) D-branes, topological objects
on which open strings end. In Refs.~\cite{dbrscatt1,dbrscatt2} the
scattering of D-particles in the type IIA string theory was studied in ten
spacetime dimensions, and it was found that the minimal size of the collision
region could be well below the ``ten-dimensional Planck length''\footnote{%
Since the analysis is in a 10-dimensional spacetime of course the relevant
length scale is the corresponding Planck length. In presence of ``large
extra dimensions'' the relation between this ten-dimensional Planck length
and the Planck length (gravitational coupling constant) we observe in the
four spacetime dimension we perceive may be nontrivial. However, the
presence of large extra dimensions is not necessary, and in fact it was not
assumed in Refs.~\cite{dbrscatt1,dbrscatt2}. Moreover, this issue related to
the possible presence of large extra dimensions is irrelevant for our line
of analysis: the quantum-gravity intuition in favour of the localization
limit $\delta X \ge L_P$, $\delta T \ge L_P$ applies equally well to the
case of a four-dimensional spacetime and to the case of a ten-dimensional
spacetime. The point we are trying to investigate is whether (for whatever
choice of number of spacetime dimensions) the localization limit $\delta X
\ge L_P$, $\delta T \ge L_P$ can be improved upon. While in studies with
different objectives it is sometimes appropriate to denote by $L_{P}^{\left(
10\right) }$ the ten-dimensional Planck length, in order to maintain a
distinction from the four-dimensional Planck length, this type of notation
is unnecessary in our analysis and we therefore denote simply by $L_P$ the
Planck length, independently of the number of dimensions chosen for
spacetime.}.

In the relevant theoretical framework the string coupling constant is small,
$g_{s} \ll 1$, and the Planck length is related to the string length, $L_s$,
by $L_{P} \sim g_{s}^{1/4}L_{s}$. One therefore has the following hierarchy
of scales:
\begin{equation}
g_{s}^{\frac{1}{3}}L_{s}\ll L_{P} \sim g_{s}^{\frac{1}{4}}L_{s}\ll
L_{s}\quad ,
\end{equation}
which will play an important role in our analysis. Also important for our
analysis of the Heisenberg-microscope procedure using D-particles is the
fact that D-particles have mass higher than the Planck mass~\cite%
{dbrscatt1,dbrscatt2}:
\begin{equation}
m=\frac{1}{g_{s}L_{s}} \sim \frac{1}{g_{s}^{3/4 } L_{P}} \gg \frac{1}{L_{P}}
\quad ,  \label{mdp}
\end{equation}
and that in a two-D-particle system, in addition to the gravitational
interaction one must take into account a sort of companion interaction (a
requirement which primarily follows from the supersymmetry of the
theoretical framework) and the two interactions largely compensate each
other, leading to a net interaction which is governed by the potential
energy
\begin{equation}
U\sim -L_{s}^{6}\frac{v^{4}}{\left( r \right) ^{7}}
+\mathcal{O}\left( \frac{v^{6}L_{s}^{10}}{\left(
r \right) ^{11}}\right) \quad ,
\label{jocU}
\end{equation}
as long as the distance $r$ between the two D-particles and the relative
velocity $v$ of the two D-particles are such that
\begin{equation}
r \gtrsim \sqrt{v}L_{s}\quad .  \label{dmin}
\end{equation}
(Although it is irrelevant for our analysis, since the localization
procedure requires a nonvanishing probe-target relative velocity, it is
noteworthy that, in particular, for a system of two D-particles at rest
there is no force.)

On the basis of these properties, the analysis of collisions between two
D-particles suggests~\cite{dbrscatt1,dbrscatt2} that the minimal dimension
of the collision region is given by $g_{s}^{1/3}L_{s}$ which is indeed below
the Planck length ($g_{s} \ll 1$
implies $g_{s}^{1/3}L_{s} \ll g_{s}^{1/4}L_{s} \sim L_P$).
In light of this result it is natural to wonder
whether D-particles can be used for accurate localization of a spacetime
point. It is from this perspective that we consider a
Heisenberg-microscope procedure in which both the probe and the target that
collide are D-particles. As in the other Heisenberg-microscope procedures we
analyzed, the $X$ coordinate can be measured by measuring where the probe
hits the ``photographic'' plate of the microscope, and the $T$ coordinate
can be measured as
\begin{equation}
T\simeq t_{i}+\frac{X}{v_{p}}\quad .
\end{equation}
And once again the uncertainties on the measurement of $X$ and $T$ are given
by (omitting again a classical-physics contribution of the type $\delta X_{c}$
which could anyway be reduced at will)
\begin{align}
\delta X& \gtrsim \delta x_{0}+d_{\min }+\delta x_{p}+\delta
x_{t} \\
\delta T& \gtrsim \delta t_{i}+\frac{\delta X}{v_{p}}+\frac{X}{v_{p}^{2}}
\delta v_{p}\quad ,
\end{align}
using notation already introduced in Eqs.~(\ref{incertezze}) and (\ref%
{incertezze2}). Here however we assume that $m_{t}=m_{p}=m$, with $m$ given
by (\ref{mdp}) (\textit{i.e.} the target and the probe are D-particles, of
mass $m$).

On the basis of the results of Refs.~\cite{dbrscatt1,dbrscatt2} we are
encouraged to aim for $g_{s}^{1/3}L_{s}$ accuracy:
\begin{equation}
\delta x^{\ast }\sim g_{s}^{\frac{1}{3}}L_{s}\quad .  \label{dxstar}
\end{equation}
and we therefore take
\begin{align}
\delta x_{0}& \simeq \delta x^{\ast }\sim g_{s}^{\frac{1}{3}}L_{s} \\
\delta v_{p}& \simeq \frac{1}{m\delta x_{0}}\sim g_{s}^{\frac{2}{3}}\quad ,
\end{align}%
where the last equation also takes into
account $p_p \simeq m v_p$ (small $v_p$).
Indeed $v_p$ must be small in order for (\ref{jocU})
to be applicable:
\begin{equation}
d_{\min }\simeq \sqrt{v_{p}}L_{s} \sim
g_{s}^{\frac{1}{3}}L_{s} \quad ~,  \label{dminvp}
\end{equation}
where we also took into account that for
consistency with our accuracy objective we must
require $d_{min} \sim g_{s}^{1/3}L_{s}$.
This leads to a choice of probe velocity of order $g_{s}^{2/3}$
\begin{equation}
v_{p}\sim g_{s}^{\frac{2}{3}}\quad .  \label{vdp}
\end{equation}

The description of $\delta x_{p}$,
\begin{equation}
\delta x_{p}\gtrsim \frac{1}{p_{p}\frac{L}{a}}\quad ,
\end{equation}
maintains the same form as in the previous Section~4 (observing again
that the momentum of the probe at early
and late times is of the same order of magnitude).

For $\delta x_{t}$ the analysis can proceed as in the previous sections but
taking into account the different form of the potential energy associated to
the probe-target interaction:
\begin{equation}
\delta x_{t}\sim \left. \frac{\Delta p_{p}\left( t\right) }{m}\right\vert
_{d_{\min }}\frac{d_{\min }}{v_{p}}\frac{L}{a}\simeq \frac{L_{s}^{6}}{p_{p}}%
\frac{v_{p}^{4}}{d_{\min }^{7}}\frac{d_{\min }}{v_{p}}\frac{L}{a}=\frac{%
L_{s}^{6}v_{p}^{2}}{md_{\min }^{6}}\frac{L}{a}\simeq \frac{1}{mv_{p}}\frac{L%
}{a}\quad ,
\end{equation}%
where we used again $p_{p}\simeq m_{p}v_{p}$  and we
estimated $[\Delta p_{p}\left( t\right) ]_{d_{min}}$ using
the same type of argument already used in the previous section.
Combining all these observations we find
that the uncertainty on the measure of $X$ can be estimated by
\begin{equation}
\delta X\gtrsim \delta x_{0}+d_{\min }+\delta x_{p}+\delta
x_{t}\gtrsim \delta x_{0}+d_{\min }+\frac{1}{mv_{p}\frac{L}{a}}+%
\frac{1}{mv_{p}}\frac{L}{a}\quad ,  \label{jonnn}
\end{equation}%
where $\delta x_{0}$ and $d_{\min }$ are of the
order of the accuracy goal $g_{s}^{1/3}L_{s}$ we are aiming for, and the
last two contributions combine to give an overall contribution which can
also be reduced to the level $g_{s}^{1/3}L_{s}$:
\begin{equation}
\min \left( \delta x_{p}+\delta x_{t}\right) =\sqrt{\frac{1}{mv_{p}\frac{L}{a%
}}\frac{1}{mv_{p}}\frac{L}{a}}=\frac{1}{mv_{p}}\sim g_{s}^{\frac{1}{3}%
}L_{s}\quad .
\end{equation}

In summary all contributions to $\delta X$ can be controlled
at the level $g_{s}^{1/3} L_{s}$, so that it is indeed legitimate to
estimate $min(\delta X) \sim g_{s}^{1/3} L_{s}$, \textit{i.e.}
we find that a space coordinate of the event of collision between
two D-particles can be measured with
better-than-Planckian accuracy.

This comes at the cost of a relatively large uncertainty on the time
coordinate of the collision event. In fact, on the basis of the observations
reported in this section, we find for $\delta T$
\begin{equation}
\delta T\gtrsim \delta t_{i}+\frac{\delta X}{v_{p}}+\frac{X\delta v_{p}}{%
v_{p}^{2}}\quad ,  \label{dtd}
\end{equation}%
where
\begin{align}
\delta t_{i}& \sim \frac{1}{\delta E_{p}}\sim \frac{E_{p}}{p_{p}\delta p_{p}}%
\sim \frac{1}{v_{p}\delta p_{p}}\sim \frac{\delta x_{0}}{v_{p}}\sim g_{s}^{-%
\frac{1}{3}}L_{s} \\
\frac{\delta X}{v_{p}}& \sim g_{s}^{-\frac{1}{3}}L_{s}\quad .
\end{align}

Also in this case (as for the other case in which we considered a
nonrelativistic probe) it is noteworthy that in a reasonable
Heisenberg-microscope setup the distance $X$ should be macroscopic, and the
contribution $Xv_{p}^{-1/2}$ should be very large. And in any case the
terms $\delta t_{i}$ and $\delta X /v_{p}$ give contributions of
order $g_{s}^{-1/3}L_{s}$, so that clearly
\begin{equation}
\delta T\gtrsim g_{s}^{-\frac{1}{3}}L_{s} \quad .  \label{fakedt}
\end{equation}
The price for the better-than-Planckian space localization ($min(\delta X)
\sim g_{s}^{\frac{1}{3}}L_{s} < g_{s}^{1/4} L_{s} \sim L_P$) is therefore a
worse-than-Planckian time localization of the event: $\delta T \gtrsim
g_{s}^{-1/3}L_{s} \gg g_{s}^{1/4}L_{s} \sim L_P$.

\section{Aside on distance measurement and the Salecker-Wigner procedure}

In our analysis of Heisenberg-microscope measurement procedures we stressed
their use in the localization of collision events. It is worth mentioning
explicitly the (rather obvious) fact that our analysis can be viewed also as
an analysis of length measurement: the coordinates $X,T$ also specify the
length $L=X$ of the interval that connects the origin of the coordinate
system and the collision event at the time $T$. Using D-particles one then
finds that, with a large time-measurement uncertainty $\delta T \gtrsim
g_{s}^{-1/3}L_{s}$, the length of the interval can me measured rather
sharply
\begin{equation}
min(\delta L) \sim g_{s}^{1/3}L_{s} \quad .  \label{fakedtbis}
\end{equation}

This can be easily verified independently using one of the distance/length
measurement procedures that are most commonly considered. In particular, some
recent studies~\cite{ng1994,gacmpla} have considered the Salecker-Wigner~%
\cite{wign} measurement procedure. The measurement of the length of, say, a
metal bar can be performed by choosing as reference time-like world line
``WL1'' one extremity of the bar, and marking by ``WL2'' the world line of
the other extremity. The length of the bar can be measured by sending a
probe of velocity $V$ at time $t=0$ from WL1 to WL2, and setting up things
in such a way that after reaching WL2 the probe is reflected back toward
WL1. By measuring the time $t=t^*$ when the probe finally returns to WL1 one
can deduce the length of the bar as $L = V t^*/2$.

This same Salecker-Wigner measurement procedure can of course also be used
to localize an event, the spacetime point marked by the arrival of the probe
at WL2, with coordinates $X=L=V t^*/2$, $T=t^*/2$.

The careful reader can easily verify, following the steps of the line of
analysis presented in the previous section, that, in the case in which both
the probe and the particle that marks the WL2 world line are D-particles, the
Salecker-Wigner measurement procedure leads to $min(\delta L) \sim
g_{s}^{1/3}L_{s}$ and $min(\delta T) \sim g_{s}^{-1/3}L_{s}$.

\section{Closing remarks}

We have argued that, as long as the localization limit is
obtained using only ordinary quantum mechanics and general relativity,
the Planck length sets absolute limits on the measurability of
spatial distances, $\delta x \ge L_P$, and time intervals, $\delta t \ge L_P$.
This is due to the interplay between
the ``Compton-wavelength uncertainty'' and
the ``Schwarzschild-radius uncertainty''.
Essentially D-particles provide us an example of the possibility
that the particles that intervene in the localization process carry
also some other charges (so that, in addition to quantum mechanics and
general relativity, some other structures come into consideration),
leading to a weakened ``Schwarzschild-radius uncertainty''.
Allowing for D-particles in the measurement
procedure, one can find $min(\delta x) \sim g_{s}^{1/12}L_P \ll L_P$
(if $g_{s} \ll 1$).

D-particles are only an example of charged probes, and with other examples
it is plausible that one could manage to further reduce $min(\delta x)$. It
therefore seems that one should not insist on simultaneous $\delta x \ge L_p$
and $\delta t \ge L_p$ uncertainties, and it is hard to say how low the
uncertainties could be in a specific quantum-gravity model. But it is
noteworthy that all of our results are consistent with the idea of a general
localization limit on the combined measurement of space and time coordinates
of a point: $\delta x \delta t \ge L_P^2$. In
the case of D-particles one finds that the sharp (better-than-Planckian)
space-position measurement can only be achieved at the cost of a rather poor
time measurement,  $\delta t \sim g_{s}^{-1/3}L_{s}$, and one ends up
finding\footnote{Some authors,
perhaps most notably Yoneya (see, \textit{e.g.}
Ref.~\cite{yoneya}), have proposed arguments in
favour of the general validity in
String Theory of an uncertainty relation $\delta x \delta t \ge L_s^2$. The
arguments adopted by Yoneya do not appear to be fully in the spirit of more
traditional measurability analyses (and therefore it would be important to
find additional evidence in support of $\delta x \delta t \ge L_s^2$ in
String Theory), but it is nonetheless noteworthy that our D-particles
analysis is consistent with this
expectation.} $\delta x \delta t \ge L_s \gg L_P^2$.
Of course in the case of localization based on massless neutral
particles one does find $\delta x \ge L_P$ and $\delta t \ge L_P$ which
results in $\delta x \delta t \ge L_P^2$.
Further investigations of the
robustness of this $\delta x \delta t \ge L_P^2$ uncertainty principle could
provide an important element of guidance for quantum-gravity research.
It may well be that the correct form of the new uncertainty principle
is somewhat different, but, in light of our analysis, it appears likely that
that it should take the form of a space/time uncertainty principle
(an uncertainty principle such that one can improve space localization
at the cost of a worse time localization of the event).

Another point that should be explored concerns the covariance of the
quantum-gravity uncertainty principles. It appears necessary~\cite{gacperspe}
to verify that the
structure of Lorentz transformations (or at least of some suitable
deformation of the Lorentz transformations~\cite{dsr}) are compatible with
the new uncertainty principle. Especially in light of this possible concern
for Lorentz covariance, it appears necessary~\cite{dopl1994} to perform more
general measurement analyses, in which all of the coordinates of the
spacetime point are considered, whereas here (and in most of the
``quantum-gravity uncertainty principle'' research) we focused on a single
space coordinate and the time coordinate.

We have also set aside here, at the level of the analysis of the
consequences of our analysis, the contributions of the type of the
term $Xv_{p}^{-1/2}$ in Eq.~(\ref{jonnn}).
These contributions appear to suggest
that the localization can be rather poor (much worse than suggested
by the $\delta x \delta t \ge L_P^2$ limit) if the spacetime point under
consideration is at a large distance from the position where the probe
starts off for the measurement procedure. This type of behaviour is expected
on the basis of some intuitions for the quantum-gravity problem, most
notably the ones that favour a role for decoherence in quantum
gravity~\cite{ng1994,gacmpla,mynapappolonpap,nggwi}.
Also this possibility deserves further investigation.

\baselineskip 12pt plus .5pt minus .5pt

\vfil

\end{document}